\documentclass[a4paper,11pt]{article}
\usepackage{jinstpub} 
\usepackage{lineno}
\usepackage{diagbox}
\usepackage{color}
\usepackage{soul}

\title{\boldmath A Configurable Ultra-Low Noise Current Source for Transition-Edge Sensor Characterization}

\author[a]{N. Li,}
\author[b]{G. Liao,}
\author[b]{D. Yan,}
\author[b]{Y. Xu,}
\author[b]{Y. Zhang,}
\author[b]{Z. Liu,}
\author[a]{S. Yuan,}
\author[b]{Y. Zhang,}
\author[b]{H. Gao,}
\author[b]{Y. Li,}
\author[b]{Y. Gu,}
\author[b]{C. Liu,}
\author[a]{H. Li,}
\author[b,1]{Z. Li
\note{Corresponding author.}}
\author[a,1]{and X. Ren}

\affiliation[a]{
Key Laboratory of Particle Physics and Particle Irradiation (MOE), Institute of Frontier and Interdisciplinary Science, Shandong University, Qingdao, Shandong, 266237, China
}
\affiliation[b]{
Key Laboratory of Particle Astrophysics, Institute of High
Energy Physics, CAS, 19B Yuquan Road, Shijingshan District,
100049, Beijing, China
}

\emailAdd{renxx@sdu.edu.cn}
\emailAdd{lizw@ihep.ac.cn}

\abstract
{
    Transition-edge sensors (TESs) are sensitive devices for detecting photons from millimeter radiation to $\gamma$ rays.
    Their photon counting efficiency and collecting area benefit from large-array multiplexing scheme, and therefore the development of multiplexing readout system has been an important topic in this field. Among the many multiplex techniques, time-division multiplexing (TDM) superconducting quantum interference device (SQUID) has been used most widely for TES readout. In this work, we design a Configurable Ultra-Low Noise Current Source (CLCS) for TES characterization and as a part of a whole TDM-TES bias control system. The CLCS is based on the feedback structure of ultra-low noise instrumentation amplifiers and low-noise, high-resolution (20 bits) digital-to-analog converter (DAC). CLCS has an ultra-high resolution of $10$ $nA$ in the $0$ to $5$ $mA$ current output range, and can perform current-voltage (IV) sweep and bias-step tests to measure key TES parameters on board. The feedback structure of the CLCS also avoids the issue of impedance mismatch.
}

\keywords{
Transition-edge sensor; Time-division multiplexing SQUID; Current bias source; Impedance matching; Low-noise level; Ultra-high resolution
}

\begin{document}
\maketitle
\flushbottom

\section{Introduction}
\label{sec:intro}
\begin{figure}[htbp]
        \centering
        \includegraphics[height=.4\textwidth]{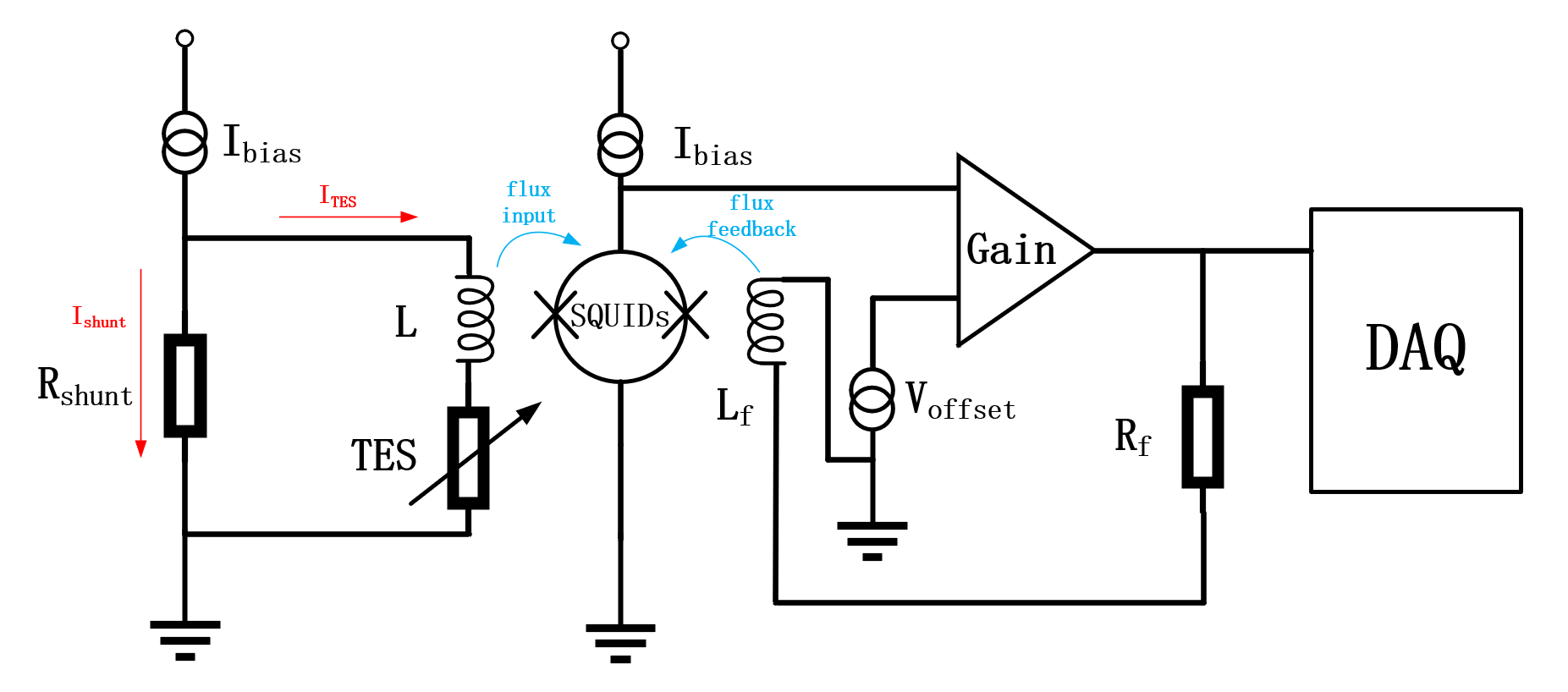}
        \caption{Schematic diagram of TES chip and simplified readout structure, which includes TES, shunt resistor $R_{shunt}$, inductive coil $L$. This is a electric thermal feedback structure with external current bias source $I_{bias}$.
        The current changes $I_{TES}$ through TES can be detected by DC-SQUIDs.
        The DC SQUID can amplify signals using its flux-lock-loop (FLL) circuit.  
        $R_f$ is the feedback resistor and $L_f$ is the feedback coil.
        \label{fig:0}}
    \end{figure}

    TESs as a type of superconducting detector have displayed high energy resolution, and and have proven their advantages in photon measurement applications ranging from millimeter radiation to $\gamma$ rays. During operation, a TES is placed on a low-temperature stage to dissipate input signal power, and is biased in an electrical circuit to turn photon signal input into electrical signal output.  Figure~\ref{fig:0} illustrates its electrical bias scheme.
    The TES is connected in parallel with a shunt resistor $R_{shunt}$ which makes the circuit a stable negative feedback loop, and in series with a inductive coil $L$ which couples TES current signal to the SQUID readout system. TES requires an external bias current $I_{bias}$ to operate~\cite{TES_in_CPD,TES_review,Takei2008,Gottardi2008,gottardi2009performance}. The TES array developed to be used with this bias and readout system is for high-energy resolution hard X-ray spectroscopy.

    We thus design a Configurable Ultra-Low Noise Current Source (CLCS) as a bias source device for our future hard X-ray TES array spectrometer with a TDM readout system. In addition to supplying low-noise DC bias current, the CLCS also integrated bias current sweeping and square-wave step-bias test functions. So that measuring TES IV curves and decay times for thermoelectric parameters can be achieved without using additional instruments such as an external function generator. This integration is potentially useful for space-limited applications, for instance, in a narrow beam-line hub or space missions.
    
    The paper is organized as follows: 
    Section~\ref{sec:sys} presents the CLCS design principles and testing system configuration;
    Section~\ref{sec:I-V} presents the thermal and electrical parameters of TES through IV curve measurement;
    Section~\ref{sec:noise} presents the noise performance of the CLCS;
    Section~\ref{sec:AC_func} presents the square wave function,
    and Section~\ref{sec:conclusion} concludes the paper.

\section{System design}
\label{sec:sys}

\subsection{Bias source electronics}
\label{sec:bias}
\begin{figure}[htbp]
        \centering
        \includegraphics[height=.3\textwidth]{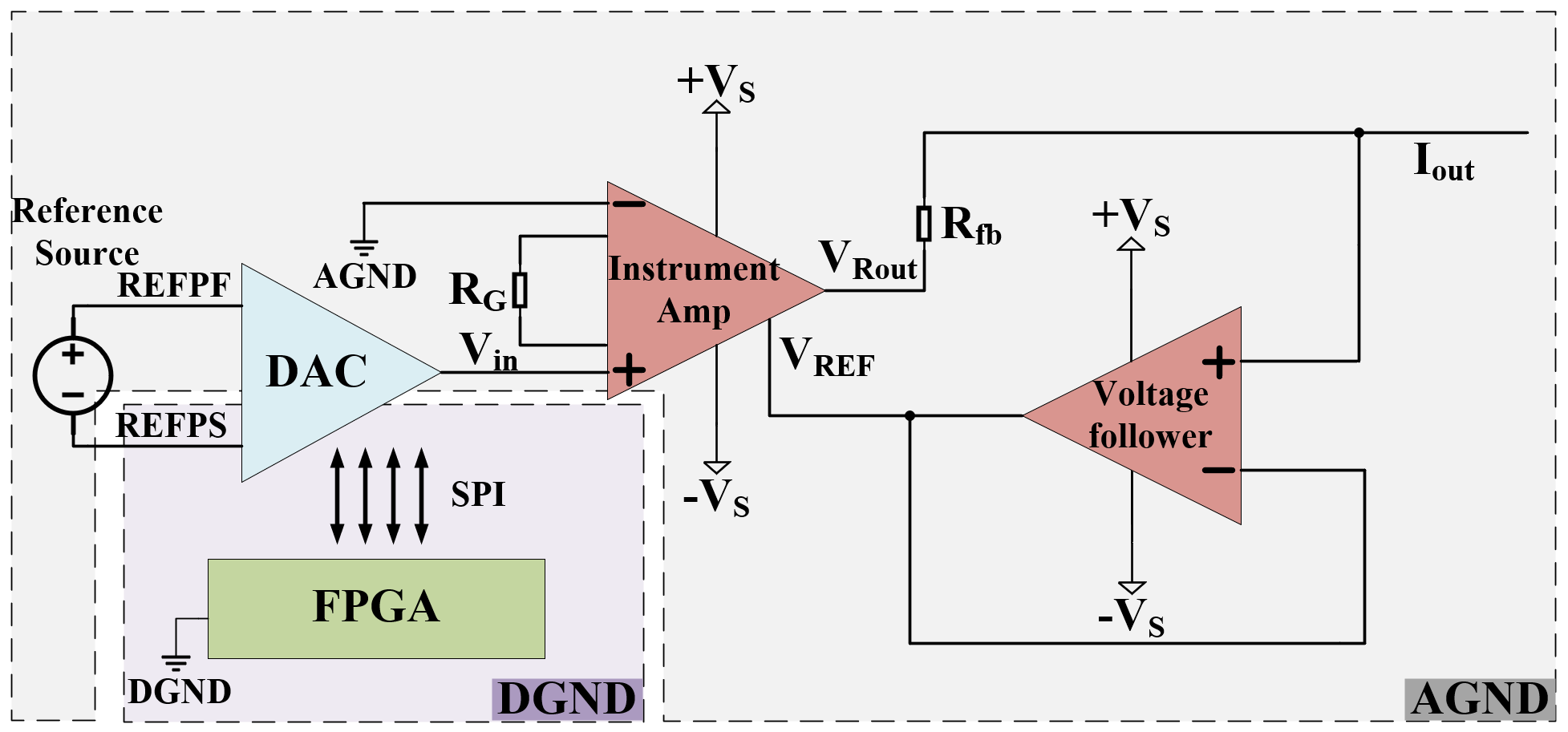}
        \qquad
        \includegraphics[height=.3\textwidth]{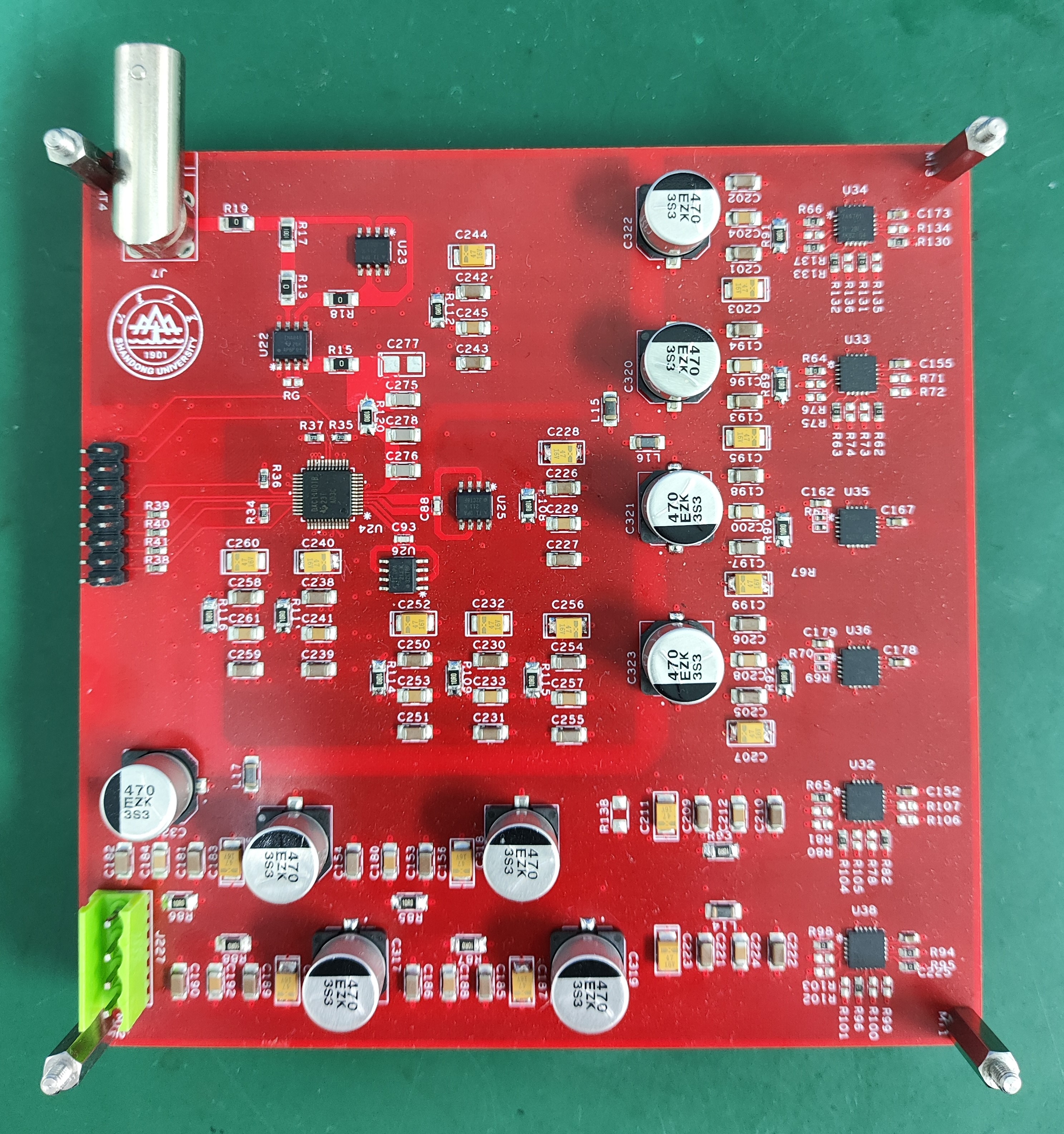}
        \caption{The left figure shows the schematic diagram of CLCS. An instrumentation amplifier and a low noise voltage follower form a feedback loop structure. DAC is used as a configable voltage source, which is controlled by FPGA. The right figure is printed circuit board (PCB) design of CLCS.\label{fig:1}}
    \end{figure}

    A common way of supplying bias current $I_{bias}$  to the TES circuit is by using a low noise source in series with a large resistor, which is usually on the order of $\sim$ 1 $k\Omega$. For example, the bias source electronics (DAC MAX5443, 16 bits) designed by NIST~\cite{TDM_1} and the Multi-Channel Electronics (MCE) designed by UBC~\cite{MCE_elec}.
    However, the impedance of the cable connecting the room temperature bias source to the cold TES stage can be several tens of ohms. Given the $\sim$ 1 $k\Omega$ series bias resistor value, the parasitic cable impedance can bring a non-negligible error in accurately calculating the $I_{bias}$ delivered to the TES circuit.
    We avoid the impedance mismatch issue by using a feedback structure~\cite{INA118, 5431858, Design_Current_Source}. 
    Thus CLCS has more wide dynamic output range.
    The bias module also contains precision instrument amplifiers with differential symmetric structure and low-noise, high-precision digital-to-analog converter (DAC).
    The operational amplifiers have high-input and low-output impedance.
    The design principle of the entire CLCS module is shown on the left side of Figure~\ref{fig:1}.
    
    To ensure high-precision regulation of current output and to enhance the driving capability of the DAC, the input $V_{in}$ of the instrumentation amplifier is connected to the output of the DAC module.
    The DAC module is DAC11001B of TI, which has 20-bit high-resolution single-channel configurable output based on the Serial Peripheral Interface (SPI) protocol.
    This high resolution enables very fine control of the TES bias point in the transition slope, and is an improvement over the widely used 16-bit devices.
    It also has low output noise, approximately $7$ $nV/\sqrt{Hz}$ as stated in the data sheet.
    The $REFPF/S$ pin of DAC11001B determines the output voltage range, which is set to $0$ to $5$ $V$ according to TES experiment requirements.
    We use INA849 of TI as the instrumentation amplifier part of CLCS. 
    Its front-end high-impedance input operational amplifier serves as a voltage follower for the DAC. 
    The internal topology ensures system stability and that the load and power supply share a common ground.

    The CLCS output current can be represented as:
    \begin{equation}
    \label{eq:1}
        \begin{aligned}
            I_{out} &= \frac{V_{Rout} - V_{REF}}{R_{fb}} \\
                    &= \frac{V_{in}Gain}{R_{fb}} \\
        \end{aligned}
    \end{equation}
    The output voltage $V_{Rout}$ of the instrumentation amplifier depends on the voltage supply limits of the device ${\pm}Vs$, which is set to be ${\pm}12$ $V$ in this case.
    $V_{REF}$ is the reference input voltage of the instrument amplifier. 
    Parasitic resistors on the $REF$ pin that is in series with the internal resistor of the instrument amplifier can disrupt the symmetry of the differential amplifier, thereby reducing the common-mode rejection ratio (CMRR). 
    Therefore, a low-noise, low-output impedance operational amplifier is used as the input for the $REF$ pin.
    Here we use the TI's low-noise operational amplifier, OPA211. 
    We connect the output port of OPA211 to its negative input port and the $REF$ port, and connect positive input to the feedback resistor $R_{fb}$, forming a voltage follower and feedback structure, as shown in Figure~\ref{fig:1}. 
    This feedback structure 
    enables a fixed voltage difference across $R_{fb}$ based on Equation \eqref{eq:1},
    thus providing a constant current.
    The value of $R_{fb}$ will affect the current source resolution and output range.
    The larger the resistance value, the higher the resolution of the current source and the smaller the output current range.
    According to the testing need of the TES, the output current range of CLCS is configured to be 0 to $5$ $mA$. By choosing a $1$ $k\Omega$ $R_{fb}$  we get a resolution of $10$ $nA$ for the bias current.
    $V_{in}$ is the input voltage of INA849, and in this single-ended design, its value is equal to the output voltage of the DAC.
    $Gain$ is the gain factor of the INA849, set to 1 in this case.

\subsection{Digital control electronics}
    To make the configuration of CLCS flexible and convenient for functional design, a Field Programmable Gate Array (FPGA) is used for digital control of CLCS.
    The model number used is Xilinx ARTIX-7 series XC7A35T and we utilize its abundant general purpose input and output (GPIO) resources. 
    The CLCS configure and monitor interactive logic is based on the SPI serial protocol.
    Joint test action group (JTAG) interface and small form-factor pluggable (SFP) gigabit optical module are used to enable information exchange between the FPGA and the host computer.

\subsection{Testing system}
\begin{figure}
      \centering
      \includegraphics[height=0.4\textwidth]{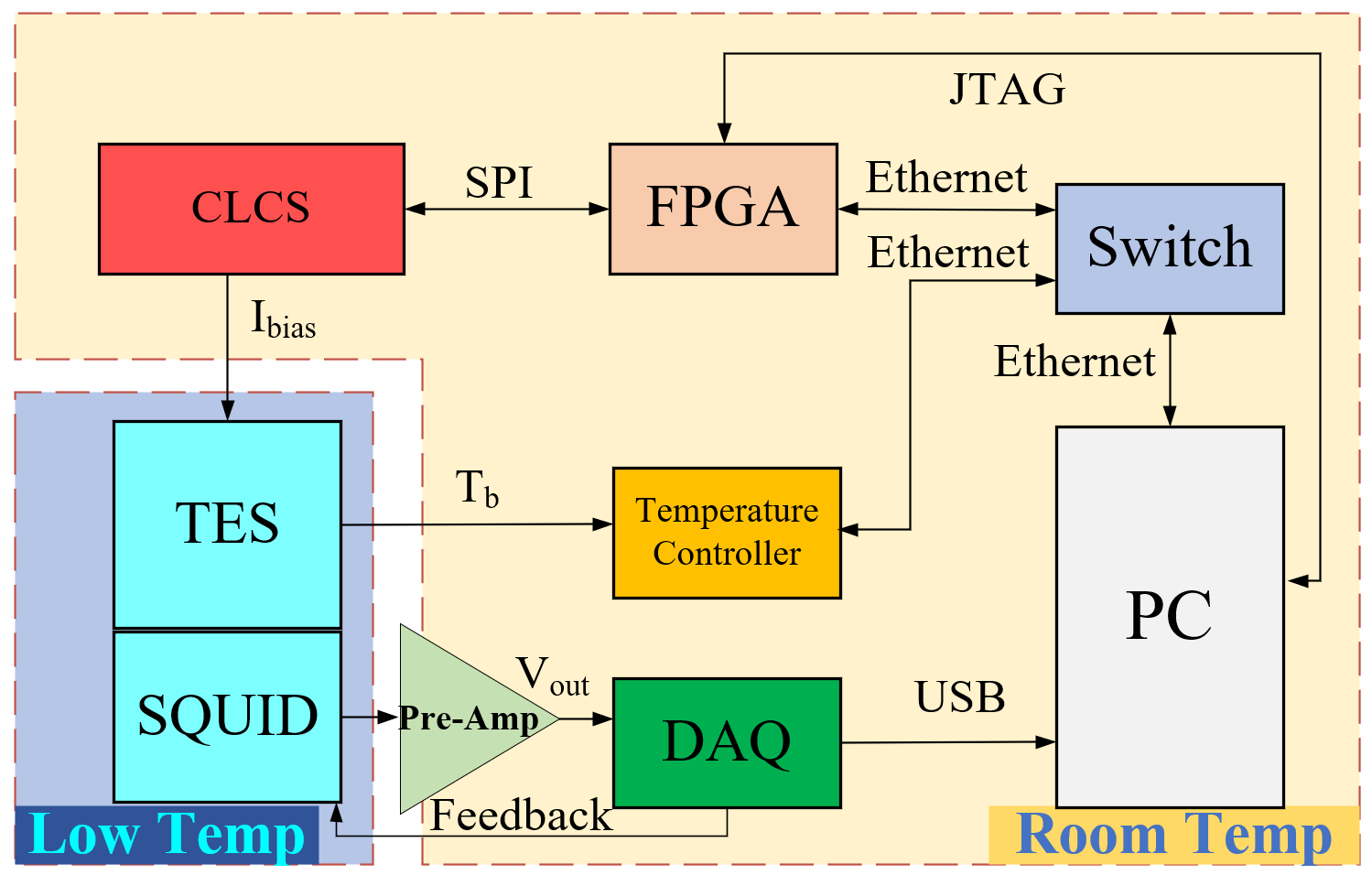}
      \qquad
      \includegraphics[height=0.4\textwidth]{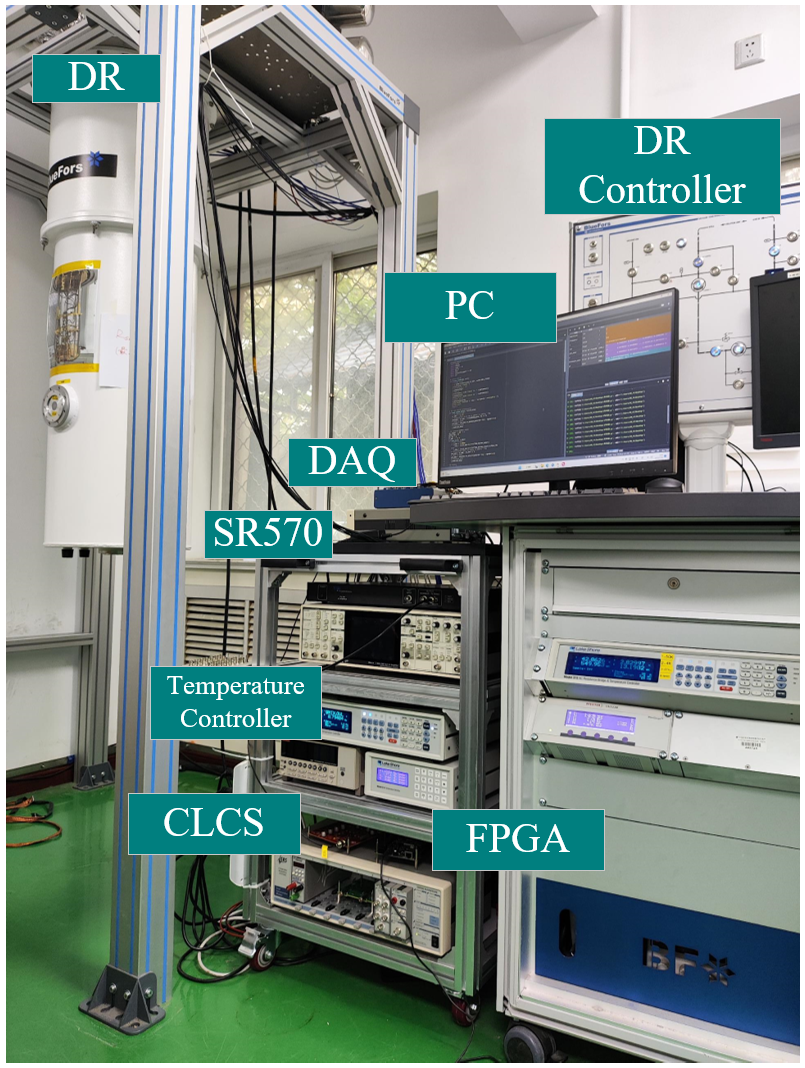}
      \caption{
      Testing system: Schematic diagram of the testing system, with the TES located in the $30$ to $60$ $mK$ temperature region, amplified by SQUID and read out by room temperature electronics (left). The control logic of the system is implemented based on Python and FPGA. Physical diagram of the system setup (right).\label{fig:2}}
    \end{figure}

    The testing system is shown in Figure~\ref{fig:2}.
    It utilizes a dilution refrigerator (DR) to operate the TES in the sub-Kelvin temperature stage. 
    The TES current is coupled to a two-stage direct current (DC) SQUIDs amplifier for low-noise readout.
    The SQUIDs amplifier is pcSQUID from STAR Cryoelectronics, which consists of two stages of SQUIDs connected in series. 
    They are operated in the flux-locked-loop (FLL) mode.
    At room temperature, the SQUIDs output signal is connected to a preamplifier for multistage amplification, and the amplified signal is then stored using a data acquisition system (DAQ).
    The DAQ system is the NI USB-6366, which is equipped with a $2$ $MSPS$ sampling rate , 16-bit resolution, and 8 channels analog-to-digital converter (ADC).
    The Lakeshore 372 temperature controller is used to monitor the real-time temperature changes of the TES cooling stage.
    The FPGA controls and monitors the CLCS output with different bias currents based on the SPI 4-wire protocol.
    The communication software framework between these devices and the host computer is implemented using the Python libraries socket, pyvisa, and nidaqmx. 
    It is based on the TCP/IP protocol or USB serial communication.

\section{IV analysis on TES samples}
\label{sec:I-V}

    The IV curve measurement under different bath temperatures $T_{bath}$ reveals several important thermal and electrical parameters about the TES. During the measurement, the voltage $V$ across the TES is swept by varying the external bias source $I_{bias}$, 
    and the TES current $I$ is sensed at the same time directly through the 2-stage SQUID system.
    Given the equilibrium condition
    \begin{equation}
    \label{eq:2}
        \begin{aligned}
            I \cdot V &= \frac{G}{nT_{c}^{n-1}}(T_{c}^{n} - T_{bath}^{n}) , \\
        \end{aligned}
    \end{equation}   
    one can fit for the TES critical temperature $T_{c}$, 
    the thermal conductance $G$ between the TES and the thermal bath, and the thermal conducting index $n$. The TES normal resistance $R_{n}$ can also be calculated from the slope of the IV curve. Furthermore, a reasonable combination of  bath temperature and DC bias current values that yield a specific bias resistance (i.e. $\%R_{n}$) can be found directly from the IV data.

\begin{figure}
      \centering
      \includegraphics[height=0.4\textwidth]{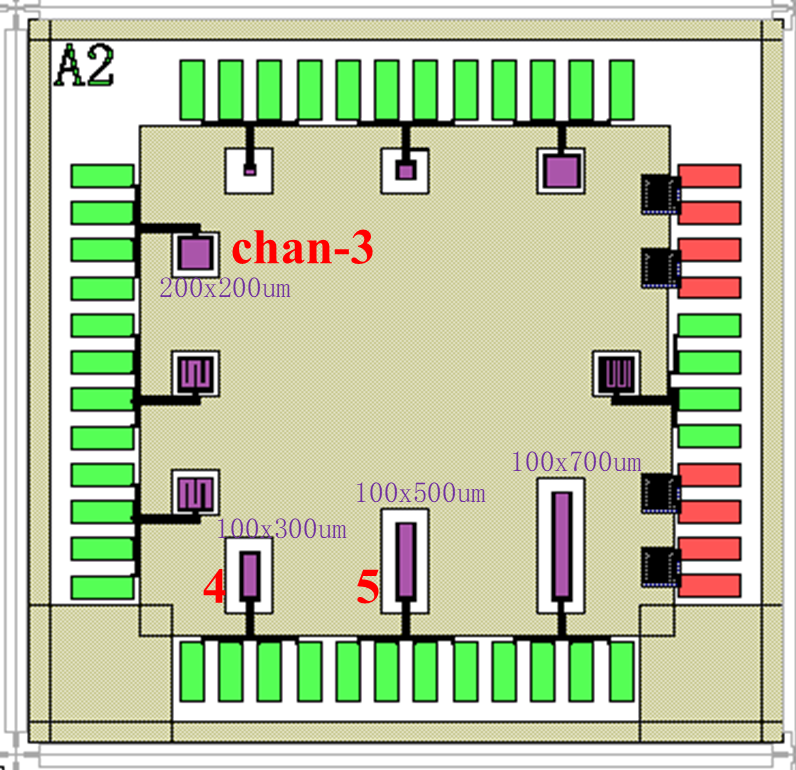}
      \qquad
      \includegraphics[height=0.4\textwidth]{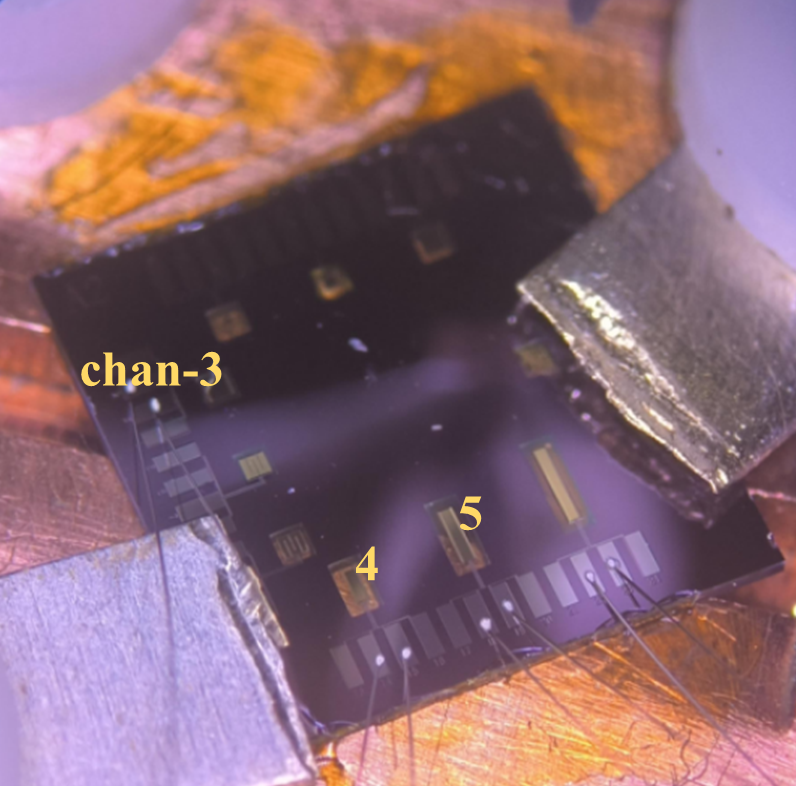}
      \caption{
      The structure and distribution of TESs: Three TESs  of different sizes labeled 3,4,5 are selected for the experiment (left). The sizes of channel 3, 4, 5 are $200 {\mu}m {\times} 200 {\mu}m$, $100 {\mu}m {\times} 300 {\mu}m$ and $100 {\mu}m {\times} 500 {\mu}m$, respectively. The expected ratio of resistance is $15:5:3$.
      Physical image shows the actual sample mounting status (right).\label{fig:3}}
    \end{figure}

    The TESs tested in this experiment are made of aluminum-manganese (AlMn) alloy and the film is deposited on $Si_{3}N_{4}$.
    This type of TESs has been widely used in long-wavelength applications in the past but rarely utilized as X-ray devices. Our goal is to make AlMn TES arrays for hard X-ray spectroscopy. As an initial step, we fabricate dark TESs of various sizes to explore the influence of the geometric factors on device thermal and electrical parameters. This is done through IV measurement with the CLCS.
    The design layout and picture of the TES devices measured are shown in figure~\ref{fig:3}.
    
    During test, the CLCS outputs a specific bias current at the command from PC through Ethernet port. 
    Meanwhile, the DAQ card and the temperature controller are synchronized to collect the SQUID output signal and temperature data.
    After each data acquisition, the system controls the bias current supply to output the next value, and repeat the above operation until the entire IV sweep is completed.
    
\begin{figure}
      \centering
      \includegraphics[width=0.3\textwidth]{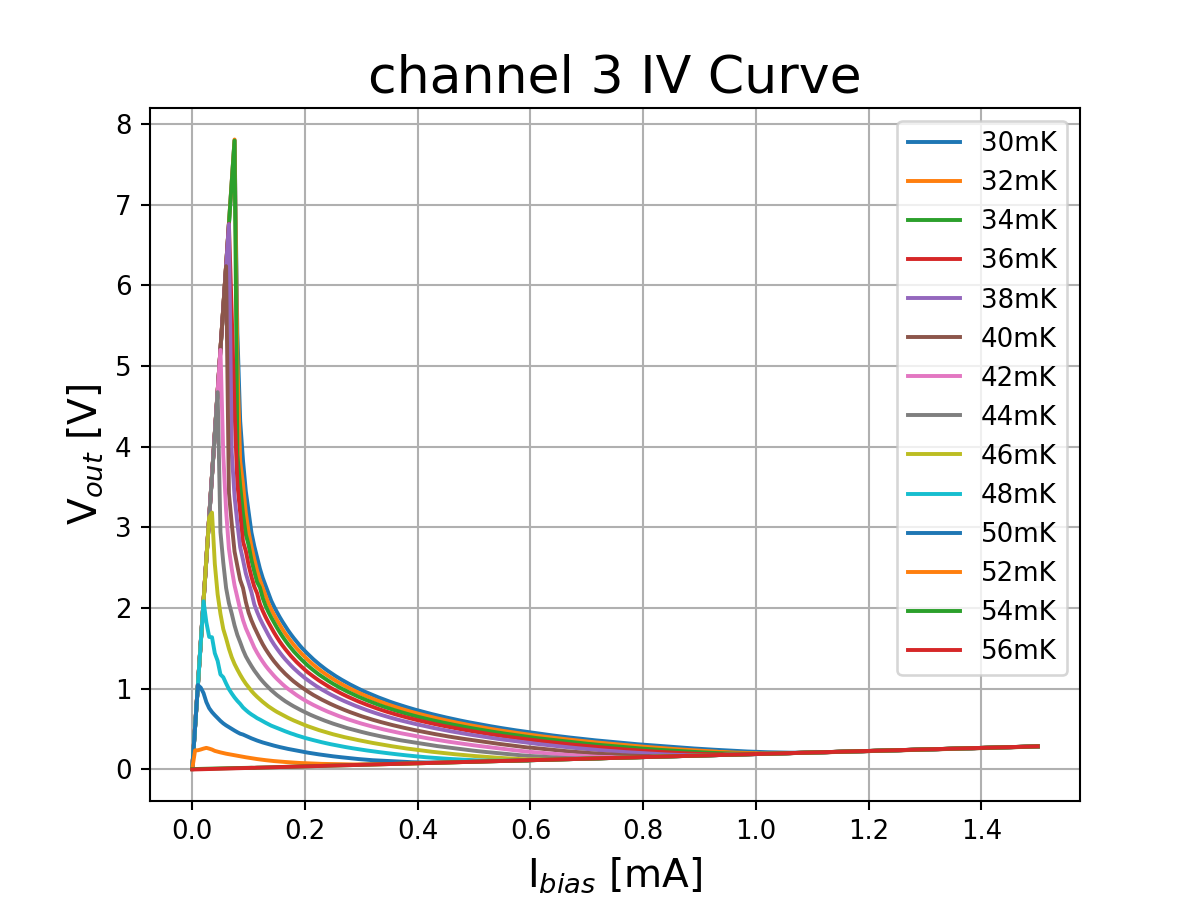}
      \includegraphics[width=0.3\textwidth]{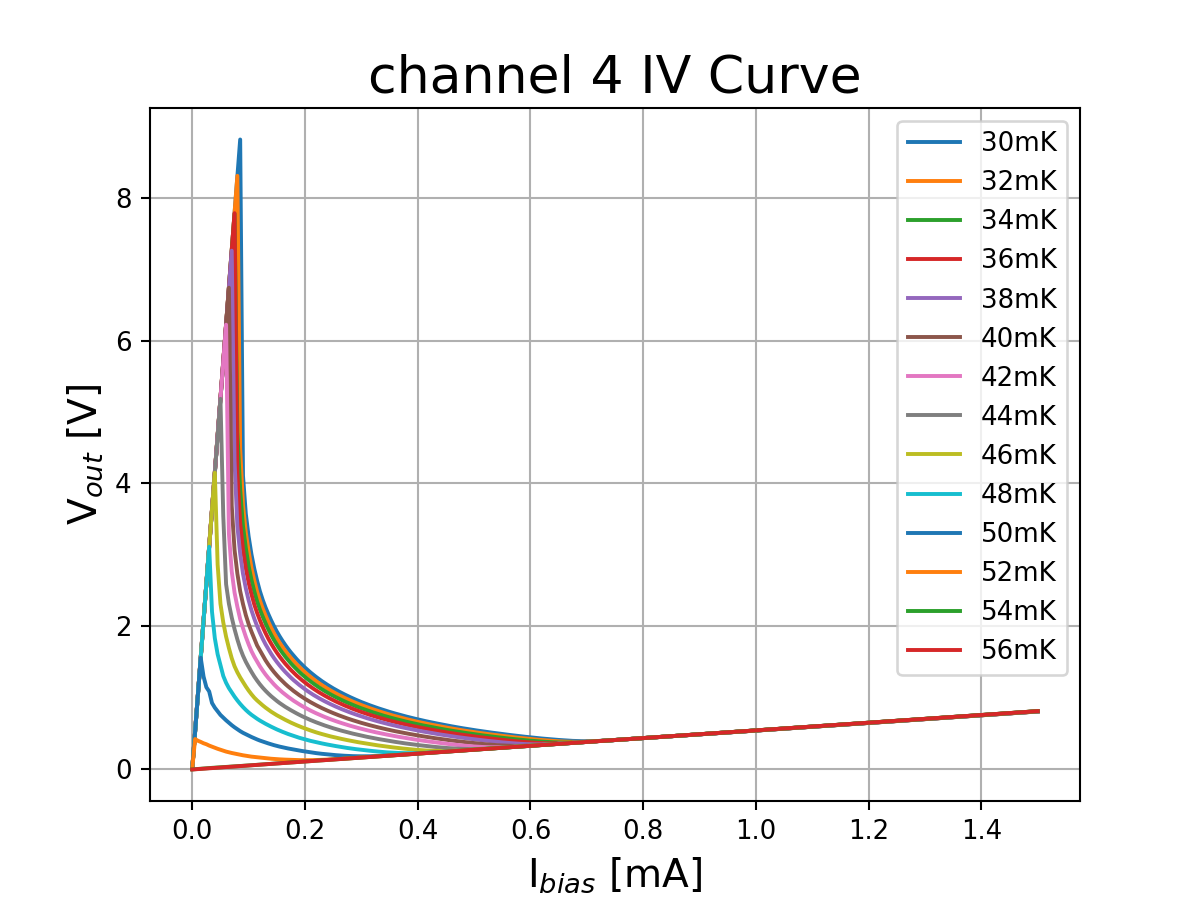}
      \includegraphics[width=0.3\textwidth]{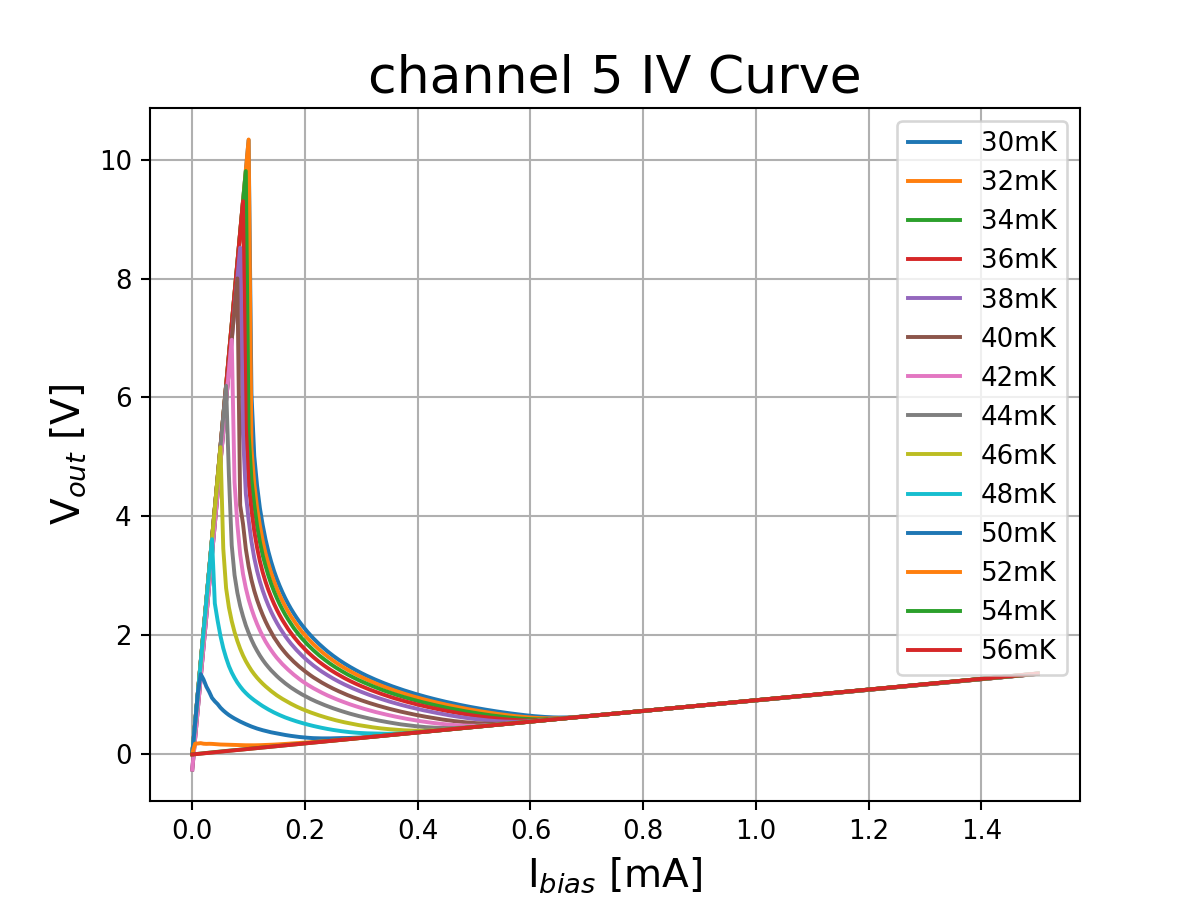}
      \caption{
        The IV test results of the three TES samples based on CLCS.
        The testing temperature ranges from $30$ $mK$ to $56$ $mK$ at every $2$ $mK$. The bias current at each temperature is swept from $1.5$ $mA$ to $0$ $mA$ with an interval of $5$ ${\mu}A$.\label{fig:4}}
    \end{figure}

    The TESs tested here are designed to have a critical temperature of $\sim$ $55$ $mK$. 
    We conduct IV tests on three different-sized TES in a temperature range of $30$ $mK$ to $56$ $mK$ at an interval of $2$ $mK$.
    To ensure that the TES goes sufficiently to the normal state, the bias current sweep starts from $1.5$ $mA$. 
    The bias step is $5$ ${\mu}A$, which expresses the high resolution performance of CLCS.
    The IV test data is shown in figure~\ref{fig:4}.
    The SQUID feedback resistor is $100$ $k\Omega$, giving a gain of $100000$ $V/A$. The TES current $I$ is calculated by dividing SQUID output voltage by the gain value, and the TES voltage $V = (I_{bias} - I)\cdot R_{shunt}$.
    Here $R_{shunt}$ is $0.25$ $m\Omega$.
    With $I$ and $V$ values at a specific bias point, in this case $80\% R_{n}$, at different bath temperatures, $G$, $n$ and $T_{c}$ can be calculated from Equation~\eqref{eq:2}.
    The results are summarized in Table~\ref{tab:1}.
    
    The ratio of $R_n$ for the three TESs is approximately $15:5:3$, which agrees well with the length-to-width ratios of the samples as shown in Figure~\ref{fig:3}. The G ratio, however, is $0.7:1:1.5$, differing obviously from the $1:1:1.5$ perimeter ratio. The deviation in $n$ of channel-3 indicates a difference in the thermal conducting mechanism. While more experiments need to be conducted to pin out the reason, one possible explanation is that the distance between the TES and the silicon substrate in channel-3 is half the distance of the other two samples. The $T_{c}$ values of these samples are about the same, and meet expectation.

    \begin{figure}[htbp]
        \centering
        \includegraphics[width=0.8\linewidth]{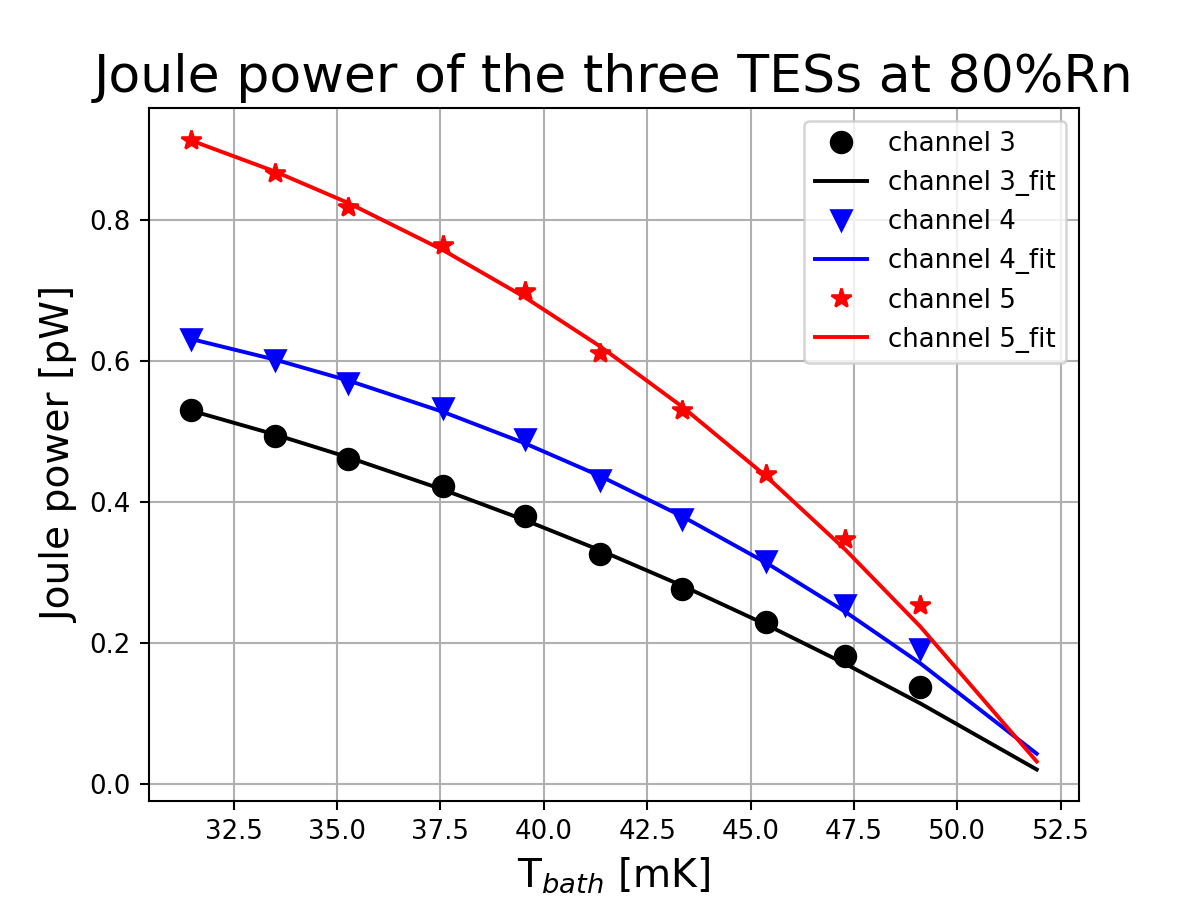}
        \quad
        \caption{The test and fit results of CLCS, for the Joule power of the three TESs. 
        \label{fig:6}}
    \end{figure}   
    \begin{table}[htbp]
        \centering
        \caption{$R_{n}$, $G$, $n$, and $T_c$ of the three TESs calculated from CLCS IV tests.\label{tab:1}}
        \begin{tabular}{c|c|c|c}
            \hline
             & $Channel{\quad}3$ & $Channel{\quad}4$ & $Channel{\quad}5$\\
            \hline
            $R_{n}{\quad}[m\Omega]$ & {$136.9$} & {$48.1$} & {$28.6$}\\
            \hline
            $G{\quad}[pW/K]$ & {$35.3$} & {$51.0$} & {$74.5$}  \\
            \hline
            $n{\quad}$ & {$2.6$} & {$3.6$} & {$3.6$}\\
            \hline
            $T_{c}{\quad}[mK]$ & {$52.5$} & {$52.8$} & {$52.4$}\\
            \hline
        \end{tabular}
    \end{table}    

\section{Noise analysis}
\label{sec:noise}
    Noise is an important factor because it determines TES energy resolution. The CLCS noise level should be sufficiently lower than TES intrinsic noise and SQUID noise.
    The noise contribution of SQUID is typically about some $pA/\sqrt{Hz}$, which is less than TES intrinsic noise.
    Our goal is to achieve $sub-pA/\sqrt{Hz}$ noise contribution from the CLCS.
    
\begin{figure}[htbp]
        \centering
        \includegraphics[width=0.8\textwidth]{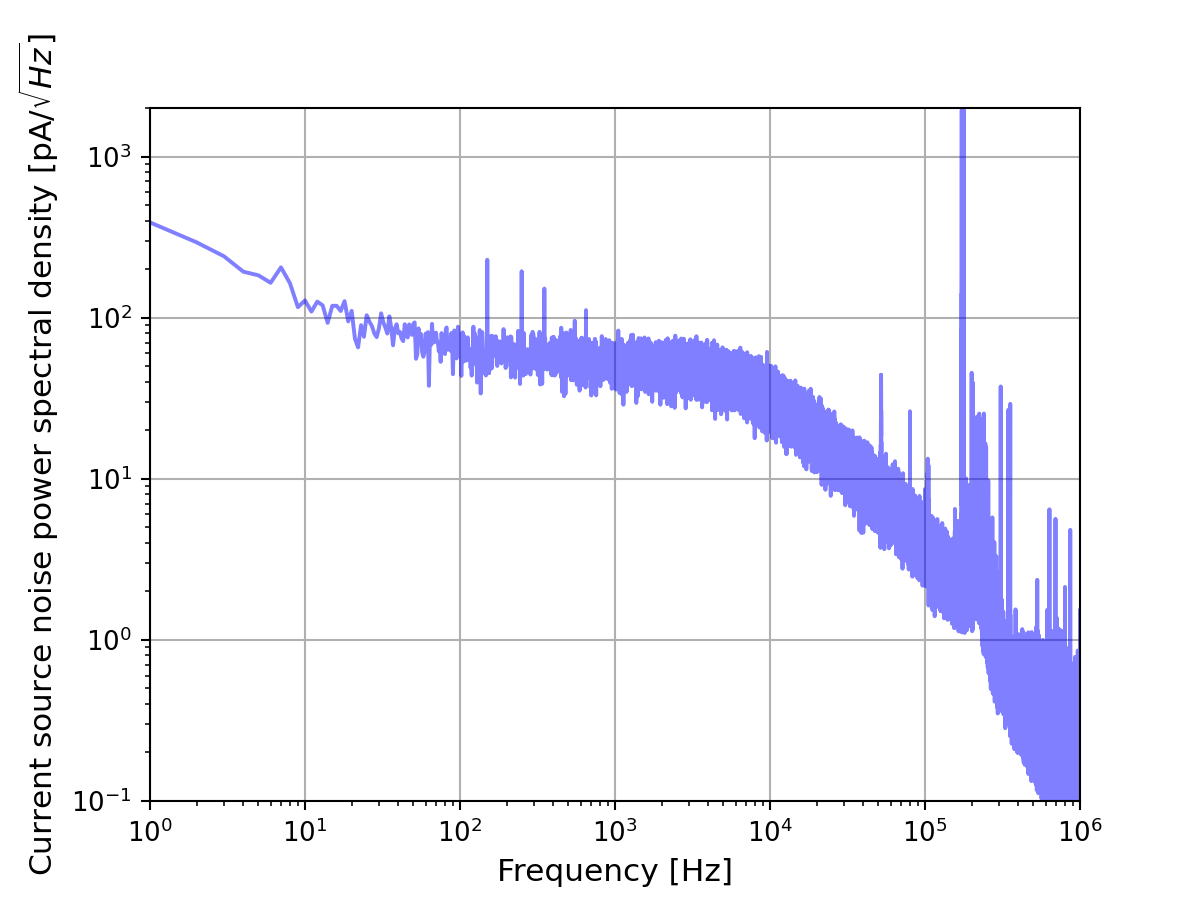}
        \caption{The intrinsic noise spectrum of CLCS. 
        The gain of SR570 is $10$ ${\mu}A/V$, and the sampling rate of DAQ is $2$ $MSPS$. Below $5$ $kHz$ ($1/f$ noise and white noise), the average noise of CLCS is about $52$ $pA/\sqrt{Hz}$, which is the mainly noise contribution of CLCS to TES system. 
        \label{fig:7}}
    \end{figure}

    By converting output current into voltage via a low-noise current amplifier (SRS-SR570), 
    the intrinsic noise of CLCS is measured and analyzed in frequency domain by fast Fourier transform function (FFT), as shown in Figure~\ref{fig:7}.
    The main noise components of CLCS includes $1/f$ noise and white noise, which is below $5$ $kHz$. We calculate the average noise value is about $52$ $pA/\sqrt{Hz}$, which can be used to calculate the noise contribution of CLCS to system.

\begin{figure}
      \centering
      \includegraphics[width=0.8\textwidth]{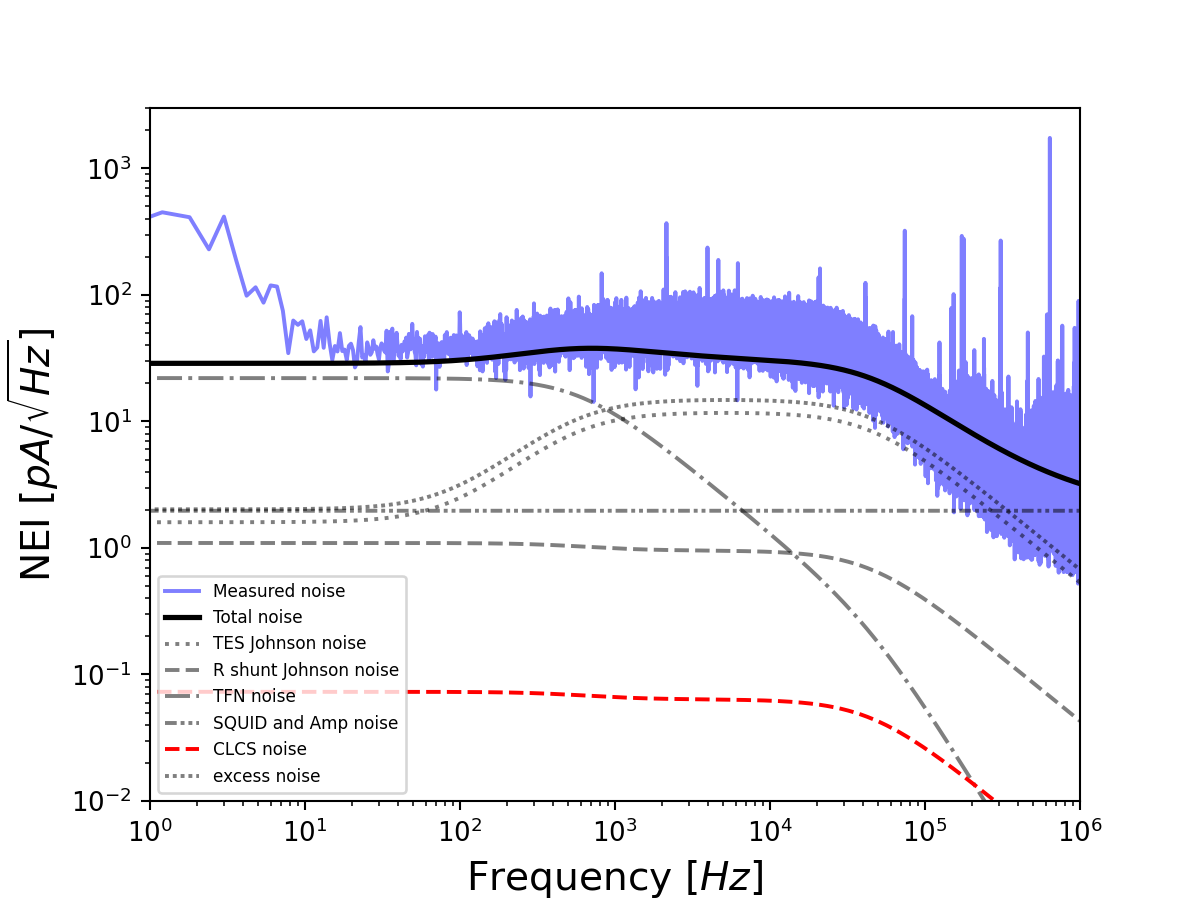}
      \caption{System noise components analysis.
      The solid blue line is the measured system noise of the channel-4 TES biased at $40\%$ $R_n$ with CLCS. 
      The solid black line is the calculated total noise. 
      The dotted gray line is the TES Johnson noise.
      The dashed gray line is the Johnson noise of shunt resistor. 
      The dash-dotted gray line is the thermal fluctuation noise (TFN) noise.
      The dash-dot-dotted gray line is the SQUID and readout electronics noise.
      The solid red line is the noise contribution of CLCS.
      The excess noise (densely dotted line) is $M$ times Johnson noise of TES, where $M=1.4$. }
      \label{fig:8}
    \end{figure}

    Figure~\ref{fig:8} shows the noise of channel-4 TES biased at $40\% R_{n}$ with CLCS. Using a one-body thermoelectric model~\cite{one-body-model}, the TES noise can be properly fitted. The low frequency regime ($<$ 1 $kHz$) is dominated by the thermal fluctuation noise (TFN), which is caused by phonon transfer between TES and thermal bath. The middle frequency regime (1 $kHz \sim$ 100 $kHz$) is dominated by TES Johnson noise, which is caused by thermal agitation of electrons in the TES. 
    There is often discrepancy between the measured and theoretical TES Johnson noise, which is called "excess noise".
    Excess noise can be expressed by $M$ times Johnson noise of TES, where $M = 0.2\sqrt{\alpha_{I}} = 1.4$~\cite{Ullom_2004} and ${\alpha}_I = {\frac{{\partial}{\log}R}{{\partial}{\log}T}}\bigg|_{I_0}$ is the unitless logarithmic temperature sensitivity of TES~\cite{TES_in_CPD} at $40\%Rn$ bias point for our TES model.
    
    In the high frequency regime ($>$ 100 $kHz$), the TES noise rolls off below SQUID and readout electronics noise. The noise contribution of CLCS is compared with all other components by multiplying it with TES circuit tranfer function:
    \begin{equation}
    \label{eq:3}
        \begin{aligned}  
            NEI_{CLCS} = \sqrt{S_{I_{CLCS}}}R_{shunt}{\mid}Y_{ext}{\mid},
        \end{aligned}
    \end{equation}
    where $NEI_{CLCS}$ is the noise equivalent current density (NEI) contribution of CLCS to TES system. $S_{I_{CLCS}}$ is the intrinsic current noise density of CLCS, which is measured and shown in Figure~\ref{fig:7}. $Y_{ext}$ is the external admittance of the TES electrothermal feedback circuit~\cite{TES_in_CPD}. 
    The noise contribution is less than $0.08$ $pA/\sqrt{Hz}$, which meets our design goal.
    Based on this analysis, the CLCS noise contribution to the whole TES system meets the requirement.

\section{AC function}  
\label{sec:AC_func}
    Count rate, which is another important factor of X-ray TESs, is directly related to the signal decay time. Furthermore, by measuring decay time at different bias currents and temperatures, the temperature sensitivity, current sensitivity and heat capacity of a TES can be fitted. The decay response can be measured by DC biasing the TES at a desired transition point while superimposing a square wave signal~\cite{cothard2020comparing}. The amplitude of the square wave should be very small compared to the DC bias that it causes negligible changes in TES bias fraction.

    This type of experiment usually is conducted with an external function generator. By implementing the alternating current (AC) function into the CLCS system, we are now able to complete the test in a more compact setup. And due to the higher current resolution of the CLCS than existing commercial function generators, we are able conduct this characterization with minimal square wave amplitude.
    
\begin{figure}
      \centering
      \includegraphics[height=0.35\textwidth]{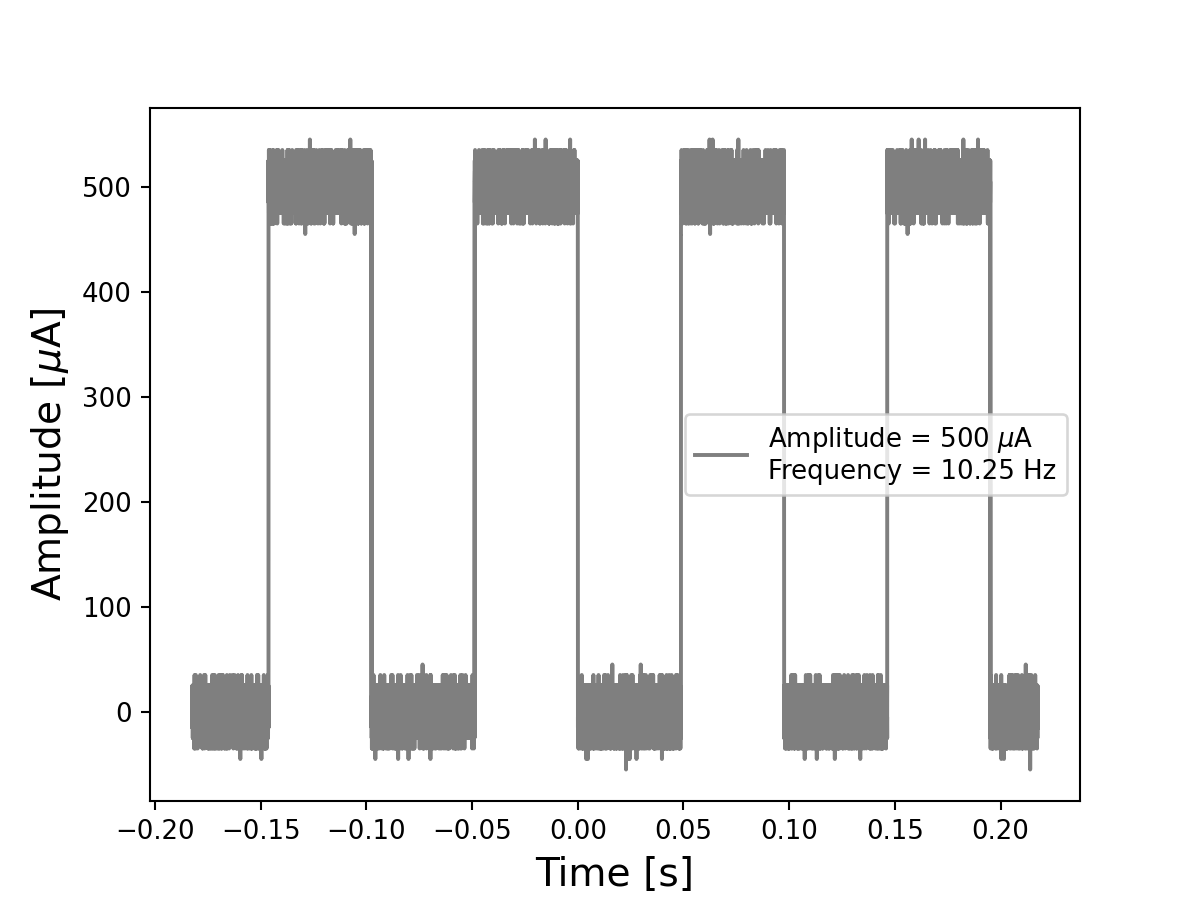}
      \qquad
      \includegraphics[height=0.35\textwidth]{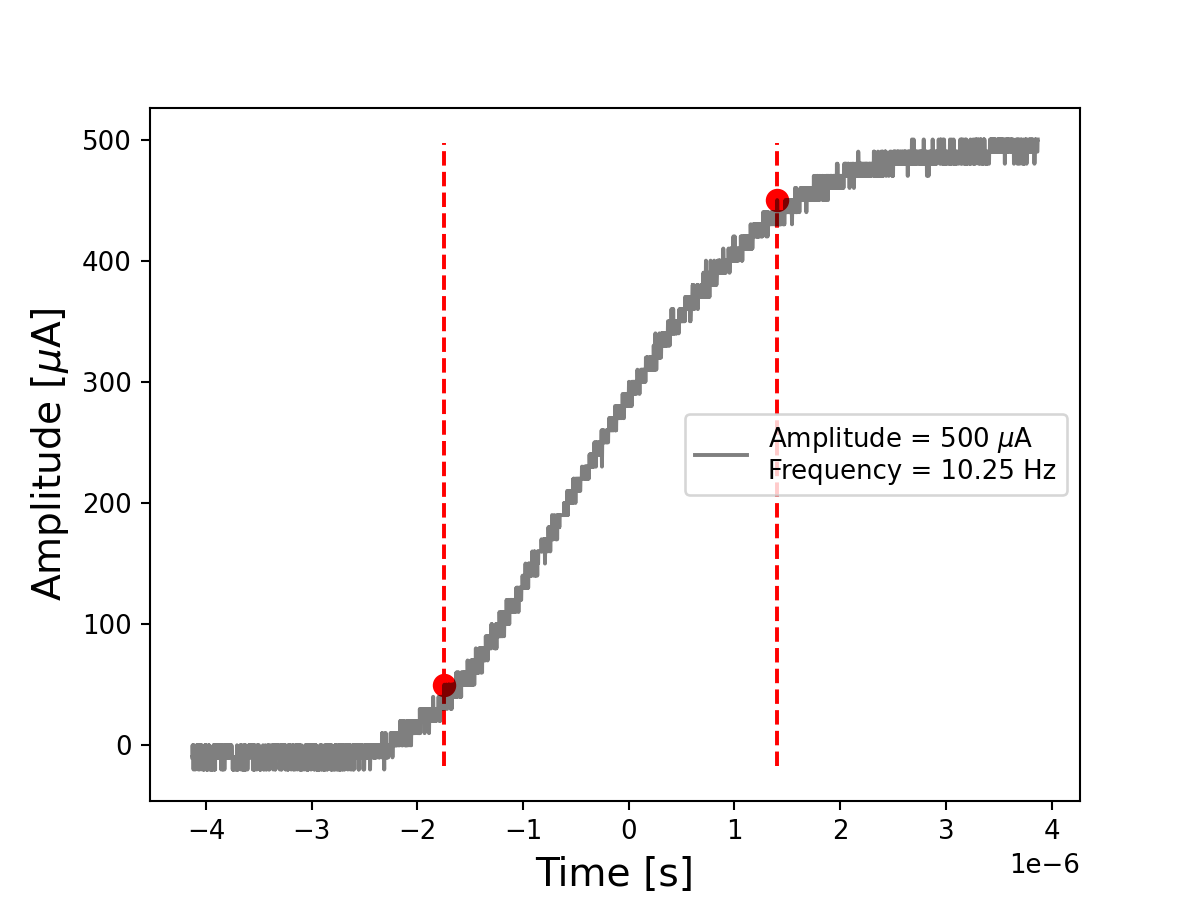}
      \caption{
      CLCS square wave test: The left figure shows a $10.25$ $Hz$ square wave with an amplitude of $500$ ${\mu}A$ generated by CLCS. The right figure is the rise time. 
      The red points and dashed lines indicate the positions of the $10\%$ and $90\%$ amplitudes. \label{fig:9}}
    \end{figure}

    For X-ray applications, the time constant of TESs is typically on the order of $\sim$ 1 $ms$. 
    In Figure~\ref{fig:9}, we show a $10.25$ $Hz$ square wave, with $3.204$ ${\mu}s$ rise time ($10\%$ to $90\%$ amplitude), generated by the CLCS, demonstrating its capability of characterizing X-ray TES decay time.  
    
    In addition to square wave, CLCS can be configured to generate sine wave up to $100$ $kHz$, which potentially can be used for complex impedance measurements. 

\section{Conclusion}
\label{sec:conclusion}
    We have designed a current source CLCS for TES bias and characterization.
    The CLCS has low noise, high resolution output and flexible configuration.
    It effectively avoids the impedance mismatch issue of voltage bias sources by applying a feedback structure.
    We performed successful IV characterization on different TES samples, demonstrated low noise contribution ($sub-pA/\sqrt{Hz}$ level), and showed AC singal generation functions with this CLCS.

    In the future work, we will upgrade the CLCS with DC-SQUID bias function and integrate it with a TDM feedback module to a whole TDM-TES control system. The final version of the control system is expected to perform TDM-TES bias, characterization, and data-acquisition tasks.


\acknowledgments
This work is supported by the National Key Research
and Development Program of China 
(Grant 
No.2022YFC2204900, 
No.2022YFC2204904, 
No.2022YFC2205000, 
No.2020YFC2201604, 
No.2022YFC2205002, 
No.2021YFC2203400,
No.2022YFC2204902,
No.2021YFC2203402,
No.2020YFC2201704).


\bibliographystyle{JHEP}
\bibliography{biblio}







\end{document}